# High Energy Polarization and the Crab: Historical Remarks

**Martin C. Weisskopf** [1]

*NASA/MSFC*
*320 Sparkman Drive, Huntsville, Al 35805, USA*
`martin@smoker.msfc.nasa.gov`

In this paper we first provide a brief history of the handful of X-ray polarization measurements that have been accomplished to date. We then outline new technological developments that indicate more detailed polarization measurements in this important band of the spectrum can, and should be, accomplished in the near future. Finally, we depart somewhat from our original assignment and mention a preliminary result of the most recent observation of the Crab with the Chandra X-Ray Observatory and compare measurement of the X-ray spectral index as a function of pulse phase with optical polarization measurements that will be presented in detail elsewhere in these proceedings.

---

[1] Speaker

## 1. The early experiments

Because of space limitations we cannot do justice to the early experiments of Robert Novick and his team at Columbia University. We give a brief summary here. For a more detailed discussion of these experiments and other X-ray polarization techniques including a discussion of the statistics, see Weisskopf et al.[1]

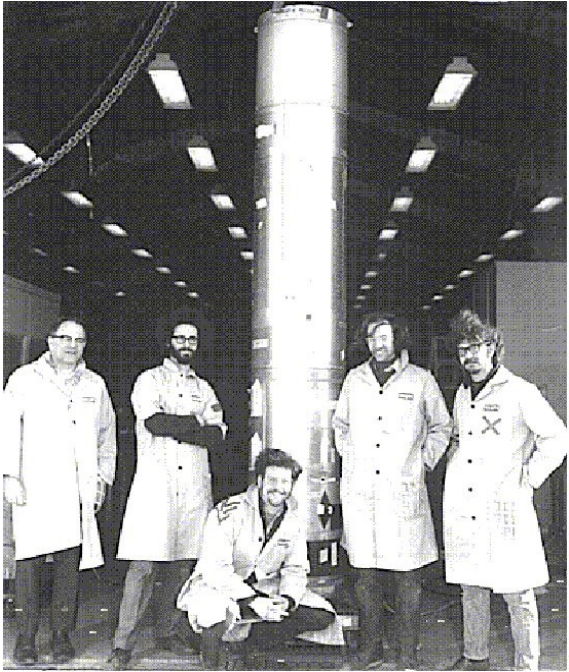

Figure 1. NASA Aerobee #1709. Left to right are R. Novick, G. Epstein, M.C. Weisskopf, R. Wolff, and R. Linke.

The first attempts to measure the polarization of the Crab Nebula system comprising the pulsar and the pulsar wind nebula were made by Bob Novick and his team at Columbia University in 1969[2]. This experiment exploited the polarization dependence of coherent and incoherent scattering utilizing blocks of lithium surrounded by proportional counters on the sides of the blocks that were at right angles to the direction of the incident flux. The experiment was launched in a sounding rocket and had only a few minutes of observation at an altitude where the Crab could be detected above the atmosphere. No polarization was detected and an upper limit of about 36% was set ($3\sigma$). In 1971 the experiment was improved to include not only a larger area lithium-scattering polarimeter, but also Bragg crystal polarimeters using "ideally imperfect" crystals of graphite. At 45 degrees incidence, where the reflection depends most sensitively on the angle of the polarization vector with respect to the plane of incidence, the graphite crystal reflects, in first order, 2.6 keV X-rays. The payload was very sophisticated for the time involving 4 doors on the sides of the rocket containing the panels of graphite crystals. These doors had to open once the rocket cleared the upper atmosphere and three of the four doors opened successfully. The crystals themselves were mounted to an approximation of a parabolic surface providing a crude level of focusing into the proportional counter detectors and hence successfully minimizing background. The payload with the doors closed and the team of scientists involved with the flight are shown in Figure 1. This flight led to the first detection of X-ray polarization of an extra-terrestrial X-ray source and found P=(15±5)% at a position angle of (156±10)° measured positive north by northeast. This success, together with Bob Novick's tenacious personality, enabled him to convince NASA to include a crystal polarimeter on the OSO-8 satellite. Observations of the Crab produced a high-precision measurement of the nebular polarization finding P=(19±1) % and a position angle of (156±2)°. This position angle is very similar to the nebula's mean position angle as measured at radio frequencies as discussed by Paolo de Bernardis elsewhere in these proceedings.

    The time resolution of the OSO-8 polarimeter also provided the ability to perform pulse-phased polarimetry, allowing one to effectively distinguish the pulsar from the nebula. The nebula was assumed to dominate the polarization during the pulse minimum. Despite the fact that the pulsar is always on, this is a reasonable assumption considering the weakness of the flux from the pulsar at pulse minimum. Unfortunately, only upper limits to the polarization of the pulsar were obtained[3] as shown in Figure 2. The contours in Figure 2 show that, at the 99%-confidence level, there is no statistically significant evidence for polarization in these data. Note, however, that the trailing edge of the interpulse appears to be 26% polarized at the 92% confidence level at a position angle of (54±17)°.

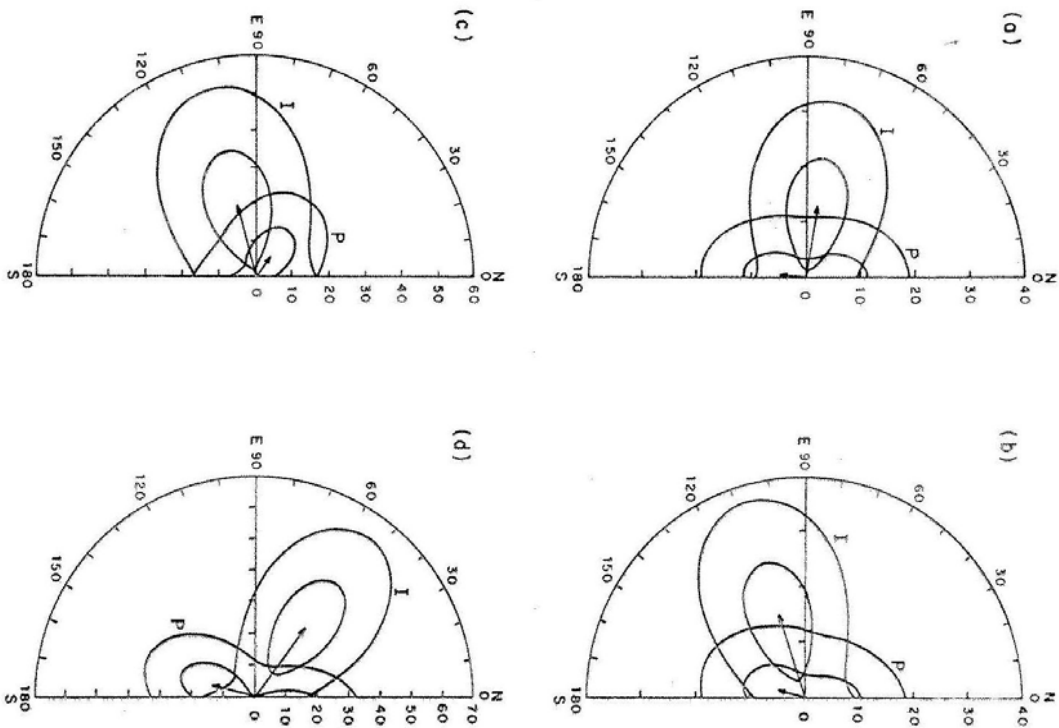

Figure 2. The polarization vectors at 2.6 keV and surrounding 67% (inner) and 99% (outer) confidence contours for the primary (P) pulse and interpulse (I) for four different ranges of pulse phase: (a) primary and interpulse together; (b) primary pulse; (c) leading edge of the primary pulse; (d) trailing edge. From Eric Silver's PhD thesis.

    These results (limits) can and should be compared with the optical polarization measurements of the pulsar first performed by Fergusen, Cocke, and Gehrels in 1974[4]. Shortly we shall see the results of more recent optical measurements by Gottfried Kanbach, Agnieszka Slowikowska, and their colleagues and discussed more completely elsewhere in these proceedings.

It is a shame that the early OSO-8 experiments were not given enough observing time[1] to perform meaningful measurements of the pulsar, as the X-Ray polarization has the distinct possibility of distinguishing among the models of the X-ray pulse production. Figure 3 illustrates three such possibilities, all of which predict more or less the same pulse shape, but can easily be distinguished by pulse-phased polarimetry.

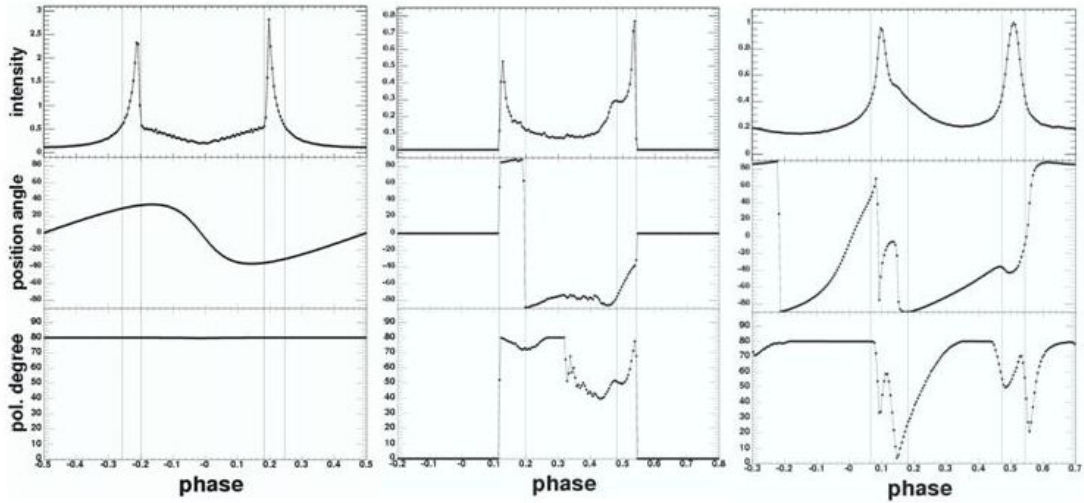

Figure 3. Three different model predictions for the Crab Pulsar intensity, polarization position angle and degree of polarization as a function of pulse phase. From left to right the models are the polar cap model ($\alpha=7°$, $\xi=12°$), the slot gap/caustic model ($\alpha=70°$, $\xi=50°$), and the outer gap model ($\alpha=65°$, $\xi=82°$). Courtesy Alice Harding.

## 2. The Future of X-Ray polarimetry

The sounding rocket and OSO-8 satellite experiments took place over 30 years ago. There are a number of reasons why there haven't been more sensitive follow up to these early exciting experiments. Polarimeters had been considered both for the Einstein and the Chandra X-ray observatories. The Einstein polarimeter became a casualty of the major descoping of the HEAO program due to problems with the Viking mission in the early 1970's. The Chandra polarimeter lost all possibility of inclusion in the mission since it was not included in the original payload and servicing was abandoned in order to get Chandra launched. Smaller satellite missions were discouraged but Bob Novick managed to get an X-Ray polarimeter placed on the original Spectrum-X mission. Having funded this effort, the community was reluctant to fund similar efforts until Spectrum-X was launched. Unfortunately, it never did. Moreover, analogous to the situation with OSO-8, the Spectrum-X polarimeter had to share a telescope's focal plane. The observing programs limited the usage to about 11 days per year, hardly adequate to perform the kind of experiments that would be needed. To my mind, these experiences point to a small, dedicated satellite mission such as the Italian POLARIX, and do not bode well for placing a polarimeter in the International X-Ray Observatory.

---

[1] A partial consequence of multiple instruments aboard the observatory, most not sharing same viewing direction.

The dedicated missions should have high sensitivity over a broad enough energy band so that meaningful astrophysics may be performed. In my view it is not sufficient to merely detect polarization, but one should be able to measure the polarization as function of energy, if for no other reason then that, as soon as one detects polarization from a particular class of X-ray source (e.g. BH accretion disc binaries), there will be a plethora of indistinguishable models explaining the detection. One will need to distinguish amongst these models, and the energy dependence may be critical. Most important will be the ability to perform spatially resolved polarimetry on physically meaningful angular scales. The Chandra image of the Crab in Figure 4 emphasizes this point, although it is by no means the only example.

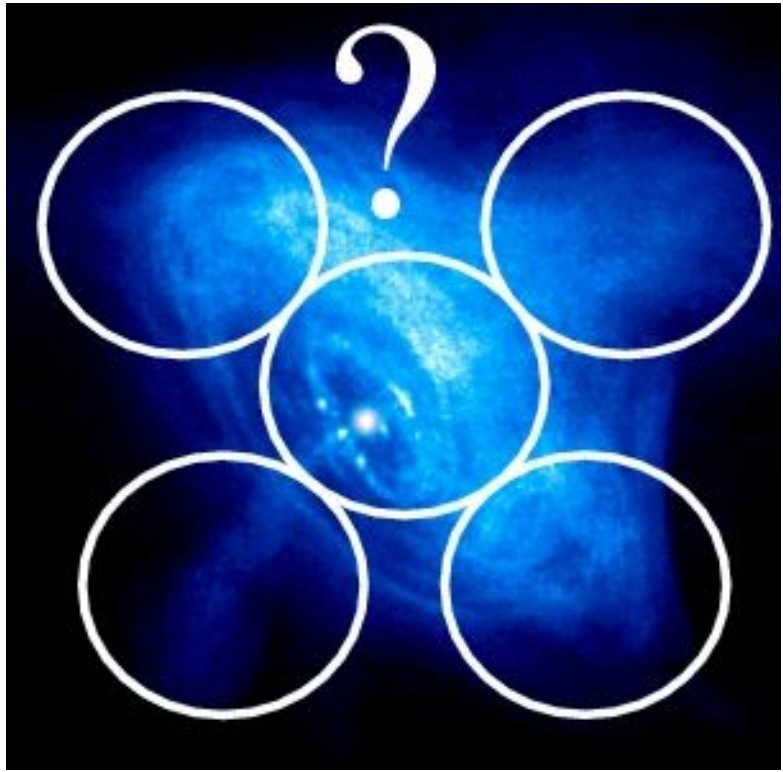

Figure 4. Chandra image of the Crab Nebula and its pulsar. The circles drawn are 30-arcsec in diameter and indicate regions which, at a minimum, one would wish to resolve to perform sensitive polarization measurements. Performing such an experiment would require a detector/telescope half-power diameter of better than 15-arcsec.

In my opinion the current detector of choice is some form of the photo-electron tracking polarimeter that makes use of the polarization dependence of the direction of the K-shell-photoelectron emitted as a result of photoelectric absorption. To my knowledge, the first electron-tracking polarimeter specifically designed to address polarimetry for X-ray astronomy was that designed by Austin and Ramsey[5] at NASA's MSFC. In their approach, the light emitted by a gaseous photo-electron-emitting material is focused and detected by a CCD camera. Another, more recent, approach was first discussed by Costa and Bellazzini and collaborators[6] wherein incident-X-ray photons enter a thin window, interact in the detector fill gas which is a mixture of low-Z ingredients designed to maximize track length and minimize diffusion. The resulting ionization track is then drifted towards a region where electron multiplication takes place and the total charge is transferred to a pixel anode array for readout. You will hear more about this detector and an application to the International X-Ray Observatory in the presentation by Enrico Costa in these proceedings. Both these versions of electron-tracking polarimeters are attractive for X-ray polarimetery as, together with an X-ray telescope, they facilitate imaging X-ray polarimetry.

## 3. Closing remarks

This conference promises to present us with a number of interesting results and intellectual challenges. It is exciting to see the results of the efforts to extract polarization data from the INTEGRAL detectors although we need to apply some caution in accepting them because of absence of a full end-to-end on-ground calibration.

Finally, as food for thought and discussion, I leave you with a *very preliminary* result based on a new Chandra observation of the Crab pulsar where we have measured the power law index as a function of pulse phase. I compare, these measurements to the recent optical polarization data I referred to above in Figure 4.

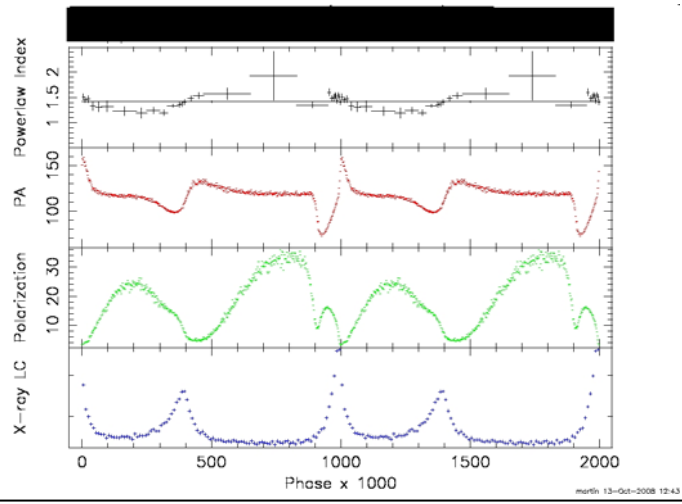

Figure 4. From top to bottom: X-Ray power law index, optical position angle, degree of optical polarization, and X-Ray flux as function of pulse phase. Optical data courtesy of Agnieszka Slowikowska.

## References

[1] M.C. Weisskopf et al. in Neutron Stars and Pulsars, W. Becker Ed. Astrophysics and Space Science Library , 2009, (357) 589

[2] R.S. Wolff, et al., *ApJ* **1970** (160) L21.

[3] E. Silver et al. *ApJ* **1978** (221) 99.

[4] D.C. Ferguson, W.J. Cocke, and T. Gehrels, *ApJ* **1974** (190) 375.

[5] R.A. Austin & B.D Ramsey, *Proc SPIE* **1992** (1943) 252

[6] E. Costa et al., *Nature* **2001** (411) 662